# About Multichannel Speech Signal Extraction and Separation Techniques

Adel Hidri[1], Souad Meddeb[2], Hamid Amiri[3]

[1,2,3] Université Tunis El Manar, Ecole Nationale d'Ingénieurs de Tunis, Signal Image and Technology of Information Laboratory BP. 37, Le Belvédère 1002 Tunis, Tunisia.
Email: [1]hidri_adel@yahoo.fr, [2]mmemeddeb@gmail.com, [3]hamidlamiri@yahoo.com

## ABSTRACT

The extraction of a desired speech signal from a noisy environment has become a challenging issue. In the recent years, the scientific community has particularly focused on multichannel techniques which are dealt with in this review. In fact, this study tries to classify these multichannel techniques into three main ones: Beamforming, Independent Component Analysis (ICA) and Time Frequency (T-F) masking. This paper also highlights their advantages and drawbacks. However these previously mentioned techniques could not afford satisfactory results. This fact leads to the idea that a combination of those techniques, which is depicted along this study, may probably provide more efficient results. Indeed, giving the fact that those approaches are still be considered as being not totally efficient, has led us to review these mentioned above in the hope that further researches will provide this domain with suitable innovations.

## 1. Introduction

Most audio signals result from the mixing of several sound sources. In many applications, there is a need to separate the multiple sources or extract a source of interest while reducing undesired interfering signals and noise. The estimated signals may then be either directly listened to or further processed, giving rise to a wide range of applications such as hearing aids, human computer interaction, surveillance, and hands-free telephony[1].

The extraction of a desired speech signal from a mixture of multiple signals is classically referred to as the "cocktail party problem" [2-3], where different conversations occur simultaneously and independently of each other.

The human auditory system shows a remarkable ability to segregate only one conversation in a highly noisy environment, such as in a cocktail party environment. However, it remains extremely challenging for machines to replicate even part of such functionalities. Despite being studied for decades, the cocktail party problem remains a scientific challenge that demands further research efforts [4].

As highlighted in some recent works [5], using a single channel is not possible to improve both intelligibility and quality of the recovered signal at the same time. Quality can be improved at the expense of sacrificing intelligibility. A way to overcome this limitation is to add some spatial information to the time/frequency information available in the single channel case. Actually, this additional information could be obtained by using two or more channel of noisy speech named multichannel.

Three techniques of MultiChannel Speech signal Separation and Extraction (MCSSE) can be defined. The first two techniques are designed to determined and over-determined mixtures (when the number of sources is smaller than or equal to the number of mixtures) and the third is designed to underdetermined mixtures (when the number of sources is larger than the number of mixtures). The former is based on two famous approaches, the Blind Source Separation (BSS) techniques [5-6-7] and the Beamforming techniques [8-9-10].

BSS aims at separating all the involved sources, by exploiting their independent statistical properties, regardless their attribution to the desired or interfering sources.

On the other hand, the Beamforming techniques, concentrate on enhancing the sum of the desired sources while treating all other signals as interfering sources. While the latter uses the knowledge of speech signal properties for separation.

One popular approach to sparsity based separation is T-F masking [11-12-13]. This approach is a special case of non-linear time-varying filtering that estimates the desired source from a mixture signal by applying a T-F mask that attenuates T-F points associated with interfering signals while preserving T-F points where the signal of interest is dominant.

In the last years, the researches in this area based their approaches on combination techniques as ICA and binary T-F masking [14], Beamforming and a time frequency binary mask [15].

This paper is concerned with a survey of the main ideas in the area of speech separation and extraction from a multiple microphones.

The following sections of this paper are organized as follows: in section 2, the problem of speech separation and extraction is formulated. In section 3, we describe some of the most techniques which have been used in MCSSE systems, such as Beamforming, ICA and T-F masking techniques. Section 4 brings to the surface the most recent methods for MCSSE systems, where combined techniques, seen previously, are used. In Section 5, the presented methods will be discussed by giving some of their advantages and limits. Finally, section 6 gives a synopsis of the whole paper and conveys some futures works.

## 2. Problem Formulation

There are many scenarios where audio mixtures can be obtained. This results in different characteristics of the sources and the mixing process that can be exploited by the separation methods. The observed spatial properties of audio signals depend on the spatial distribution of a sound source, the sound scene acoustics, the distance between the source and the microphones, and the directivity of the microphones.

In general, the problem of MCSSE is stated to be the process of estimating the signals from $N$ unobserved sources, given from $M$ microphones, which arises when the signals from the $N$ unobserved sources are linearly mixed together as presented in Figure 1.

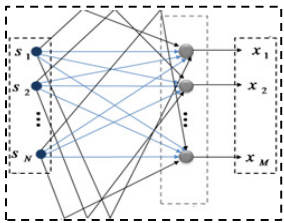

**Figure1. Multichannel Problem Formulation.**

The signal recorded at the $j^{th}$ microphone can be modeled as:

$$x_j(n) = \sum_{i=1}^{N}\sum_{p=1}^{P} h_{ji}^p S_i(n - \tau_{ji}^p) \quad j = 1 \dots M \quad (1)$$

Where $S_i$ and $x_j$ are the source and mixture signals respectively, $h_{ji}$ is a P-point Room Impulse Response (RIR) from source $i$ to microphone $j$, $P$ is the number of paths between each source-microphone pair and $\tau$ is the delay of the $p^{th}$ path from source $j$ to microphone $i$ [9-14]. This model is the most natural mixing model, encountered in live recordings called echoic mixtures.

In free-reverberation environments (p = 1), the samples of each source signal can arrive at the microphones only from the line of sight path, and the attenuation and delay of source $i$ would be determined by the physical position of the source relative to the microphones. This model, called anechoic mixing, is described by the following equation obtained from the previous equation:

$$x_j(n) = \sum_{i=1}^{N} h_{ji} S_i(n - \tau_{ji}) \quad j = 1..M \quad (2)$$

The instantaneous mixing model is a specific case of the anechoic mixing model where the samples of each source arrive at the microphones at the same time ($\tau_{ji} = 0$) with differing attenuations, each element of the mixing matrix $h_{ji}$ is a scalar that represents the amplitude scaling between source $i$ and microphone $j$. From the equation (2), instantaneous mixing model can be expressed as:

$$x_j(n) = \sum_{i=1}^{N} h_{ji} S_i(n), \quad j = 1..M \quad (3)$$

## 3. MCSSE Techniques

### 3.1. Beamforming Technique

Beamforming is a class of algorithms for multichannel signal processing. The term Beamforming refers to the design of a spatio-temporal filter which operates on the outputs of the microphone array [8]. This spatial filter can be expressed in terms of dependence upon angle and frequency. Beamforming is accomplished by filtering the microphone signals and combining the outputs to extract (by constructive combining) the desired signal and reject (by destructive combining) interfering signals according to their spatial location [9].

Beamforming for broadband signals like speech can, in general, be performed in the time domain or frequency domain. In time domain Beamforming, a Finite Impulse

Response (FIR) filter is applied to each microphone signal, and the filter outputs combined to form the Beamformer output. Beamforming can be performed by computing multichannel filters whose output is ŝ(t) an estimate of the desired source signal as shown in Figure 2.

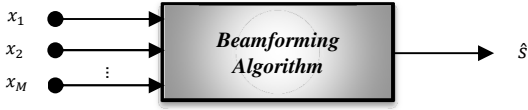

**Figure 2. MCSSE with Beamforming Technique.**

The output can be expressed as:

$$\hat{s}(t) = \sum_{i=1}^{N}\sum_{p=0}^{P-1} w_{i,p} x_i(t-p) \qquad (4)$$

Where *P-1* is the number of delays in each of the *N* filters.

In frequency domain Beamforming, the microphone signal is separated into narrowband frequency bins using a Short-Time Fourier Transform (STFT), and the data in each frequency bin is processed separately.

Beamforming techniques can be broadly classified as being either data-independent or data-dependent. Data independent or deterministic Beamformers are so named because their filters do not depend on the microphone signals and are chosen to approximate a desired response. Conversely, data-dependent or statistically optimum Beamforming techniques are been so called because their filters are based on the statistics of the arriving data to optimize some function that makes the Beamformer optimum in some sense.

### 3.1.1. Deterministic Beamformer

The filters in a deterministic Beamformer do not depend on the microphone signals and are chosen to approximate a desired response. For example, we may wish to receive any signal arriving from a certain direction, in which case the desired response is unity over at that direction. As another example, we may know that there is interference operating at a certain frequency and arriving from a certain direction, in which case the desired response at that frequency and direction is zero. The simplest deterministic Beamforming technique is delay-and-sum Beamforming, where the signals at the microphones are delayed and then summed in order to combine the signal arriving from the direction of the desired source coherently, expecting that the interference components arriving from off the desired direction cancel to a certain extent by destructive combining. The delay-and-sum Beamformer as shown in Figure 3 is simple in its implementation and provides easy steering of the beam towards the desired source. Assuming that the broadband signal can be decomposed into narrowband frequency bins, the delays can be approximated by phase shifts in each frequency band.

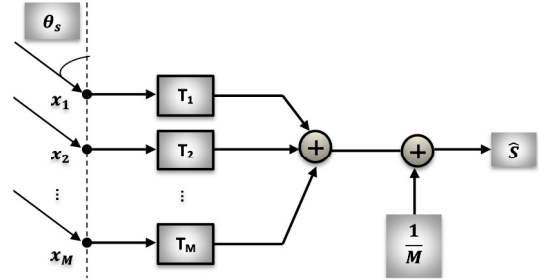

**Figure 3. Delay-and-sum Beamforming.**

The performance of the delay-and-sum Beamformer in reverberant environments is often insufficient. A more general processing model is the filter-and-sum Beamformer as shown in Figure 4 where, before summation, each microphone signal is filtered with FIR filters of order *M*. This structure, designed for multipath environments namely reverberant enclosures, replaces the simpler delay compensator with a matched filter. It is one of the simplest Beamforming techniques but still gives a very good performance.

As it has been shown that the deterministic Beamformer is far from being fully manipulated independently from the microphone signals, the statistically optimal Beamformer is tightly linked and tied to the statistical properties of the received signals.

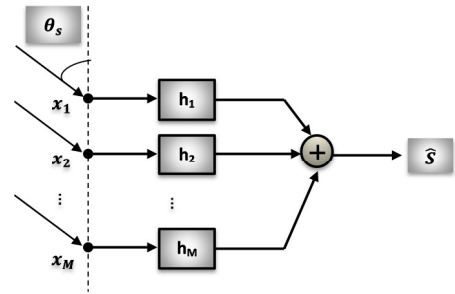

**Figure 4. Filter and sum Beamforming.**

### 3.1.2. Statistically optimum Beamformer

Statistically optimal Beamformers are designed basing on the statistical properties of the desired and interference signals. In this category, the filters designs are based on the statistics of the arriving data to optimize some function that makes the Beamformer optimum in some sense. Several criteria can be applied in the design of the Beamformer, e.g., maximum signal-to-noise ratio (MSNR), minimum mean-squared error (MMSE), minimum variance distortionless response (MVDR) and linear constraint minimum variance (LCMV). A summary of several design criteria can be found in [10]. In general, they aim at enhancing the desired signals, while rejecting the interfering signals.

Figure 5 depicts the block diagram of Frost Beamformer or an adaptive filter-and-sum Beamformer as proposed in [16], where the filter coefficients are adapted using a constrained version of the Least Mean-Square (LMS) algorithm. The LMS is used to minimize the noise power at the output while maintaining a constraint on the filter response in look direction. Frost's algorithm belongs to a class of LCMV Beamformers.

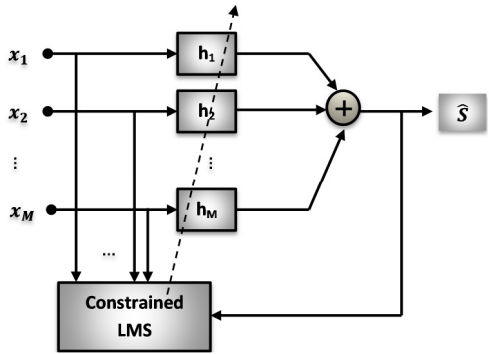

**Figure 5. Frost Beamformer.**

In an MVDR Beamformer [17], the power of the output signal is minimized under the constraint that signals arriving from the assumed direction of the desired speech source are processed without distortion.

An improved solution to the constrained adaptive Beamforming problem decomposes the adaptive filter-and-sum Beamformer into a fixed Beamformer and an adaptive multi-channel noise canceller. The resulting system is termed the Generalized Side-lobe Canceller (GSC) [18], a block diagram of which is shown in Figure 6. Here, the constraint of a non distorted response in look direction is established by the fixed Beamformer while the noise canceller can then be adapted without a constraint.

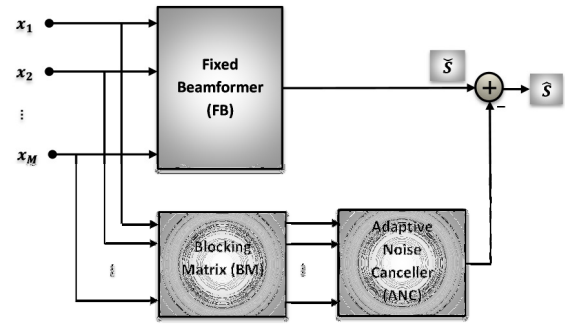

**Figure 6. GSC Beamformer.**

The fixed Beamformer can be implemented via one of the previously discussed methods, for example, as a delay-and-sum Beamformer. To avoid distortions of the desired signal, the input to the Adaptive Noise Canceller (ANC) must not contain the desired signal. Therefore, a Blocking Matrix (BM) is employed such that the noise signals are free of the desired signal. The ANC then estimates the noise components at the output of the fixed Beamformer and subtracts the estimate. Since both the fixed Beamformer and the multi-channel noise canceller might delay their respective input signals, a delay in the signal path is required. In practice, the GSC can cause a degree of distortion to the desired signal, due to a phenomenon known as signal leakage. Signal leakage occurs when the BM fails to remove the entire desired signal from the lower noise cancelling path. This can be particularly problematic for broad-band signals, such as speech, as it is difficult to ensure perfect signal cancellation across a broad frequency range. In reverberant environments, it is in general difficult to prevent the desired speech signal from leaking into the noise cancellation branch.

In practice, the basic filter-sum Beamformer seldom exhibits the level of improvement that the theory promises and further enhancement is desirable. One method of improving the system performance is to add a post-filter to the output of the Beamformer. In [19], a multichannel Wiener filter (MWF) technique, which is depicted in Figure 7, was proposed. The MWF produces an MMSE estimate of the desired speech component in one of the microphone signals, hence simultaneously performing noise reduction and limiting speech distortion. In addition, the MWF is able to take speech distortion into account in its optimization criterion, resulting in the speech distortion weighted multichannel Wiener filter (SDW-MWF) [20].

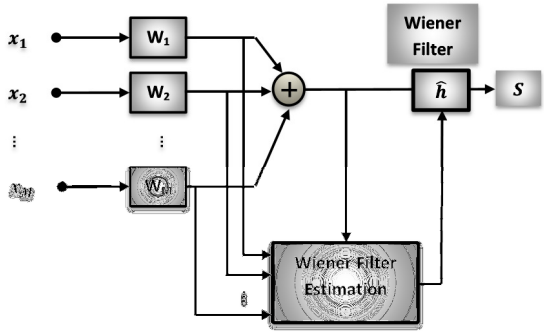

**Figure 7. Filter and sum Beamformer with post-filter.**

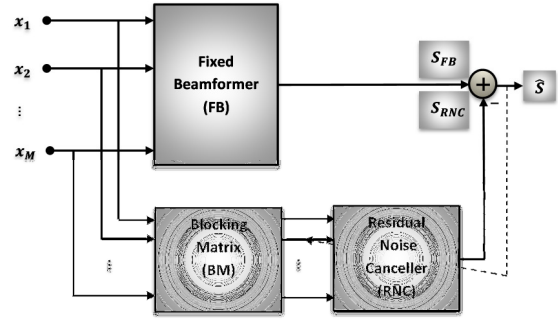

**Figure 8. LCMV Beamformer and RNC.**

Several researchers have proposed modifications to the MVDR for dealing with multiple linear constraints, denoted LCMV. Their works were motivated by the desire to apply further control to the array Beamformer beam-pattern, beyond that of a steer-direction gain constraint. Hence, the LCMV can be applied to construct a beam-pattern satisfying certain constraints for a set of directions, while minimizing the array response in all other directions.

In [8], Shmulik Markovich presented a method for source extraction based on the LCMV Beamformer. This Beamformer has the same structure of GSC but there is sharp difference between both of them. While the purpose of the ANC in the GSC structure is to eliminate the stationary noise passing through the BM, in the proposed structure the Residual Noise Canceller (RNC) is only responsible for the residual noise reduction as all signals, including the stationary directional noise signal, are treated by the LCMV Beamformer. It is worthy to note that the role of the RNC block is to enhance the robustness of the algorithm.

However the LCMV Beamformer was designed to satisfy two sets of linear constraints. One set is dedicated to maintain the desired signals, while the other set is chosen to mitigate both the stationary and non-stationary interferences. A block diagram of this Beamformer is depicted in Figure 8. The LCMV Beamformer comprises three blocks: the fixed Beamformer responsible for the alignment of the desired source and the BM blocks the directional signals. The output of the BM is then processed by the RNC filters for further reduction of the residual interference signals at the output. For more details concerning each block of this Beamformer and for the various definitions of the constraints see [8].

### 3.2. Independent Component Analysis Technique

Another approach to source separation and extraction is to exploit statistical properties of source signals. One popular assumption is that the different sources are statistically independent, and is termed ICA [21]. In ICA, separation is performed on the assumption that the source signals are statistically independent, and does not require information on microphone array configuration or the direction of arrival (DOA) of the source signals to be available. The procedure of ICA technique is shown in figure 9:

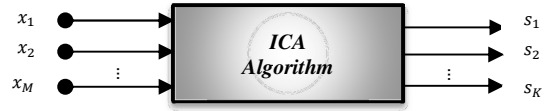

**Figure 9. MCSSE with ICA technique.**

In the instantaneous and determined mixtures case, the source separation problem can be performed by estimating the mixing matrix $A$, and this allows one to compute a separating matrix $W = A^{-1}$ whose output:

$$\hat{s}(t) = A^{-1}x(t) = W\,x(t) \qquad (5)$$

$\hat{s}(t)$ is an estimate of the source signals. The mixing matrix $A$ or the separating matrix $W$ is determined so that the estimated source signals are as independent as possible. The separating matrix functions as a linear spatial filter or Beamformer that attenuates the interfering signals.

ICA can then be applied to separate the convolutive mixtures either in the time domain [6-22], in the transform domain [6-7-23], or their hybrid [12-13].

The time-domain approaches attempt to extend instantaneous ICA methods for the convolutive case. Upon convergence, these algorithms can achieve good separation performance due to the accurate measurement of statistical independence between the segregated signals.

However, the computational cost associated with the estimation of the filter coefficients for the convolution operation can be very demanding, especially when dealing with reverberant (or convolutive) mixtures using filters with long time delays.

To reduce computational complexity, the frequency domain approaches [7] transform the time-domain convolutive model into a number of complex-valued instantaneous ICA problems, using the Short-Time Fourier Transform (STFT). Many well-established instantaneous ICA algorithms can then be applied at each frequency bin. Nevertheless, an important issue associated with this approach is the so-called permutation problem, i.e. the permutation of the source components at each frequency bin may not be consistent with each other. As a result, the estimated source signals in the time domain (using an inverse STFT) may still contain the interferences from the other sources due to the inconsistent permutations across the frequency bands. Different methods have been developed to solve the permutation problem [24]. Most methods for resolving frequency-dependent permutation fall into one of three categories: those that exploit specific signal properties of the Discrete Fourier Transform (DFT), those that exploit specific properties of speech [25] and those that exploit specific geometric properties of the sensor array, such as directions of arrival [26]. All three classes of methods require additional information about the measurement setup or the signals being separated.

Hybrid time–frequency methods tend to exploit the advantages of both time and frequency domain approaches, and considers the combination of the two types of methods. In particular, the coefficients of the FIR filter are typically updated in the frequency domain and the nonlinear functions are adopted in the time domain for evaluating the degree of independence between the source signals. In this case, no permutation problem exists any more, as the independence of the source signals is evaluated in the time domain.

Nevertheless, a limitation with the hybrid approaches is the increased computational load induced by the back and forth movement between the two domains at each iteration using the DFT and inverse DFT.

### 3.3. T-F Masking Technique

When the number of sources is greater than the number of microphones, linear source separation using the inverse of the mixing matrix is not possible. Hence, ICA cannot be used for this case. Here the sparseness of speech sources is very useful and time–frequency diversity plays a key role [27]. However, under certain assumptions, it is possible to extract a larger number of sources. Sparseness of a signal means that only a small number of the source components differ significantly from zero.

One popular approach to sparsity-based separation is T-F masking [13]. This approach is a special case of non-linear time-varying filtering that estimates the desired source from a mixture signal by applying a T-F mask. It attenuates T-F points associated with interfering signals while preserving T-F points where the signal of interest is dominant.

With the binary mask approach, we assume that signals are sufficiently sparse, and therefore, assumptions could be built that at most one source is dominant at each time–frequency point. If the sparseness assumption holds, and if an anechoic situation can be possibly assumed then the geometrical information about the dominant source at each time–frequency point can be estimated. The geometrical information is estimated by using the level and phase differences between observations. Taking into consideration this information for all time–frequency points, the points can be grouped into N clusters. Giving the fact that an individual cluster corresponds to an individual source, therefore a separation of each signal is obtained by selecting the observation signal at time–frequency points in each cluster with a binary mask. The best known approach may be the Degenerate Unmixing Estimation Technique (DUET) [28], which can separate any number of sources using only two mixtures. The method is valid when sources are W-disjoint orthogonal [29], that is, when the supports of the windowed Fourier transform of the signals in the mixture are disjoint. For anechoic mixtures of attenuated and delayed sources, the method allows its users to estimate the mixing parameters by clustering relative attenuation-delay pairs extracted from the ratios of the T-F representations of the mixtures. The estimates of the mixing parameters are then used to partition the T-F representation of one mixture to recover the original sources.

Figure 10 shows the flow of the binary mask approach, where the separation procedure [30] is formulated by the next five steps:

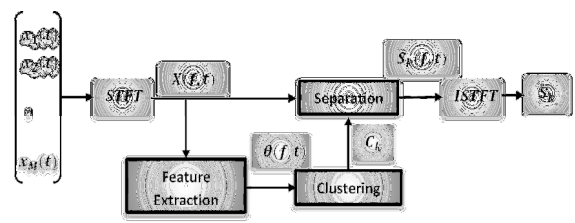

**Figure 10. Blok diagram of MCSSE with T-F masking.**

- **STEP 1**: T-F domain transformation: The binary mask approach often uses a T-F domain representation. First, time-domain signals $x_i(t)$ sampled at frequency $f_s$ are transformed into frequency domain time series signals $X_i(f,t)$ with a T-point STFT:

- **STEP 2:** Feature extraction: The separation can be achieved by gathering the T-F points where just one signal is estimated to be dominant only if the sources are sufficiently sparse. To estimate such T-F points, some features $\theta(f,t)$ are calculated by using the frequency domain observation signals $X(f,t)$. Most existing methods use the level ratio and/or phase difference between two observations as their features $\theta(f,t)$.

- **STEP 3:** Clustering: this step is concerned with the clustering of the features $\theta(f,t)$ where each cluster corresponds to an individual source. With an appropriate clustering algorithm, the features $\theta(f,t)$ are grouped into *N* clusters *C1…CN*, where *N* is the number of possible sources. To name one of the many existing clustering algorithms: the k-means clustering algorithm [31].

- **STEP 4:** Separation: based on the clustering result, the separated signals $\hat{s}_k(f,t)$ are estimated. Here a T-F domain binary mask which extracts the T-F points of each cluster has to be designed as:

$$M_k(f,t) = \begin{cases} 1 & \theta(f,t) \in C_k \\ 0 & \text{otherwise} \end{cases} \quad (6)$$

The separated signals can be expressed as:

$$\hat{s}_k(f,t) = M_k(f,t)X_j(f,t) \quad (7)$$

where *j* is a selected sensor index.

- **STEP 5:** The reconstruction of separated signal: an inverse STFT (ISTFT) and the overlap-and-add method are finally used to obtain the outputs $\hat{s}_k(t)$.

## 4. Combination Techniques

Some proposed methods have efficient separation results in a real cocktail party environment. In the recent years, researchers resorted to methods based on the combination techniques as viewed previously. Two MCSSE systems of combination techniques are presented in this section. The first is based on the combination of ICA and T-F masking [14-32]. The second is based on Beamforming and T-F masking [15].

### 4.1. ICA and binary T-F masking

In [32], ICA is applied to separate two signals by using two microphones. Based on the ICA outputs, T-F masks are estimated and a mask is applied to each of the ICA outputs in order to improve the Signal to Noise Ratio (SNR). This method is applicable to both instantaneous and convolutive mixtures. The performance of this method is compared to the DUET algorithm [28]. The result of this comparison proposes that the method in [32] produces better results for instantaneous mixtures and comparable results for convolutive mixtures.

In the same way, the paper [15] suggested two-microphone approach to separate convolutive speech mixtures. This approach is based on the combination of ICA and ideal binary mask (IBM), together with a post-filtering process in the cepstral domain. The convolutive mixtures are first separated using a constrained convolutive ICA algorithm. The separated sources are then used to estimate the IBM, which are further applied to the T-F representation of original mixtures. IBM is a recent technique, originated from computational auditory scene analysis (CASA) [33]. It has shown promising properties in suppressing interference and improving quality of target speech. IBM is usually obtained by comparing the T-F representations of target speech and background interference, with 1 assigned to a T-F unit where the target energy is stronger than the interference energy and 0 otherwise. In order to reduce the musical noise induced by T-F masking, cepstral smoothing is applied to the estimated IBM. The segregated speech signals are observed to have considerably improved quality and limited musical noise. The performance of this method is compared with the algorithm in [32]. The results of this comparison show that this method is faster than the proposed in [32]. Although the results for SNR are comparable, this method outperforms significantly the method in [34] in terms of computational efficiency. Although the mentioned methods [15-32] which combine both the ICA and T-F masking techniques have contributed to the advancement of this area of research, they still have some deficiencies. Indeed, the limitations appear in two different conditions. The first can be detected when those proposed algorithms are applied to the underdetermined cases. The second is when those approaches are put into action in highly reverberant speech mixtures.

### 4.2. Beamforming and T-F masking

In [15], J. Cemark and al. proposed a MCSSE system from convolutive mixtures in three stages employing T-F binary masking (TFBM), Beamforming and a non-linear post processing technique. TFBM was exploited as a pre-separation process and the final separation was accomplished by multiple Beamformers. His method removes the musical noise and suppresses the interference in all T-F slots. A block diagram of his proposed three-stage system is shown in Figure 11[15].

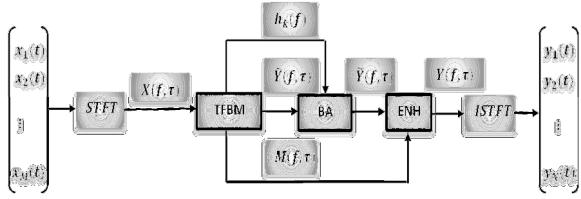

**Figure 11. System Block Diagram.**

After STFT, a TFBM is used to estimate the mixing vector $\hat{h}_k(f)$ and T-F mask $M_k(f,\tau)$ so that the pre-separated signal:

$$\hat{y}_k(f,\tau) = [\hat{y}_{1k}(f,\tau),\ldots,\hat{y}_{Mk}(f,\tau)]^T$$
$$= M_k(f,\tau)X(f,\tau) \quad (8)$$

becomes an estimate of the source

$$\hat{y}_k(f,\tau) \approx \hat{h}_k(f,\tau)s_k(f,\tau) \quad (9)$$

$M_k(f,\tau)$ extracts the T-F slots of cluster $C_k$ whose members are estimated to belong to the source signal $s_k(f,\tau)$.

$$M_k(f,\tau) = \begin{cases} 1 & X(f,\tau) \in C_k \\ 0 & \text{otherwise} \end{cases} \quad (10)$$

The cluster $C_k$ can be estimated by using for example DUET [28].

The Beamforming Array (BA) is depicted as using $D$ Beamformers to estimate the $k^{th}$ target signal $y_k(f,\tau)$. The ultimate objective of the BA is to compose D different mixtures from the pre-separated signals provided by TFBM, which are later filtered by D Beamformers. All these Beamformers are designed to enhance the desired signal. Each input mixture includes the pre-separated target signal and different pre-separated jammers. The major issue is that all the jammers must be used at least once. As a result of Beamforming, the enhanced target signal D times is gotten. By the end of the process, all the outputs of the Beamformers are gathered together.

The third stage is devoted to the enhancement (ENH). The enhancement improves the interference suppression in the T-F slots of the desired signal $\tilde{y}_k(f,\tau)$ where $M_k(f,\tau) = 0$.

Finally, the vector of the separated target signals $y(f,\tau) = [y_1(f,\tau),\ldots,y_N(f,\tau)]^T$ is transformed back into the time domain by ISTFT.

This system provides high separation performance. It had shown that a BA eliminates the musical noise caused by conventional TFBM. Furthermore, the interference in the extracted T-F slots of the desired signal is minimized. The third stage of this system permits to control the level of musical noise and interference in the output signal.

In [9], Dmour and al. proposed an MCSSE algorithm combines T-F masking techniques and mixture of Beamformers. This system is composed of two major stages. In the first stage, the mixture T-F points are partitioned into a sufficient number of clusters using one of the T-F masking techniques. In the second stage, they use the clusters which are dealt with in the first stage to calculate covariance matrices. These covariance matrices and the T-F masks are then used in the mixture of MPDR Beamformers. The resulting non-linear Beamformer has low computational complexity and eliminates the musical noise results from T-F masked outputs at the expense of lower interference attenuation. The mixture of MPDR Beamformers can be viewed as a post-processing step for sources separated by T-F masking. The contribution of those methods [15-9] is beyond any doubt, but they still have some areas of weaknesses. Those shortcomings are obvious at the level of these approaches applications. Actually, the methods have to adopt two main stages which render the whole process more complex in its implementation. They are also limited once applied in a highly reverberant environment.

## 5. Discussion

The difficulty of source separation and extraction depends on the number of sources, the number of microphones and their arrangements, the noise level, the way the source signals are mixed within the environment, and on the prior information about the sources, microphones, and mixing parameters. A vast number of methods have been found in order to come out with practical solutions to the problem of MCSSE. Those methods can be categorized, in this paper, into three main techniques named: Beamforming, ICA and T-F masking.

Beamforming techniques are applied to microphone arrays with the aim of separating or extracting sources and improving intelligibility by means of spatial filtering. Despite the fact that they have many additions to this field of research they still have some limitations to name but a few: the non-stationarity of speech signals, the multipath propagation in real environments and the underdetermined cases (when the sources outnumbered the microphones). Given those shortcomings which go against a better fulfillment of these techniques, it is clear that using the Beamforming approach only is obviously insufficient and does not convey flawless results in specific circumstances.

ICA technique is performed on the assumption that the source signals are statistically independent, and does not require information on microphone array configuration or the DOA of the source signals to be available. It has been studied extensively, the separation performance of developed algorithms is still limited, and leaves much room for further improvement. This is especially true when dealing with reverberant and noisy mixtures. For example, in the frequency-domain approaches, if the frame length for computing the STFT is long and the number of samples within each window is small, the independence assumption may not hold any more. On the other hand, a short size of the STFT frame may not be adequate to cover the room reverberation, especially for mixtures with long reverberations for which a long frame size is usually required for keeping the permutations consistent across the frequency bands. Taking into consideration these flaws which handicap a better fulfillment of this technique, it is safe to argue that using the ICA approach only is clearly insufficient and coveys restricted results.

When the number of sources surpasses the number of microphones, linear source separation using the inverse of the mixing matrix is not possible. As a result, ICA cannot be used for this case. Here the sparseness of speech sources is very practical and T-F diversity plays a crucial role. However, under certain suppositions, it is possible to extract a larger number of sources. The assumption that the sources have a sparse representation under an adequate transform is a very popular assumption. The T-F mask techniques seem versatile; however, separated signals with a T-F mask usually contain a non-linear distortion that is called the musical noise.

Few methods used aforementioned techniques, proposed in the literature, have satisfactory separation results in a real cocktail party environment. Based on the pros and cons of the multichannel techniques, researchers resort to methods relying on the combination techniques [28-35-36].

## 6. Conclusion

Separating desired speaker signals from their mixture is one of the most challenging research topics in speech signal processing. Indeed, it is very crucial to be able to separate or extract a desired speech signal from noisy observations. Actually, researchers who tended to use the single channel method found it –to a certain extent- limited and unable to offer more efficiency. This explains the recent inclination towards the use of the multichannel method which gives more flexibility and tangible results. Three basic techniques of multichannel algorithms are presented in this paper: Beamforming, ICA and T-F masking. However, despite of the existence of the vast number of applied algorithms using those three fundamental techniques mentioned previously, no reliable results have been achieved. This shortcoming leads automatically to the thought that a probable combination may offer better ends. What is worth mentioning is that a human has a remarkable ability to focus on a specific speaker in that case. This selective listening capability is partially attributed to binaural hearing. Two ears work as a Beamformer which enables directive listening, then the brain analyzes the received signals to extract sources of interest from the background, just as blind source separation does. Based on this principle, we hope to separate or extract the desired speech by combining Beamforming and blind source separation.